\documentclass{evn2004}
\usepackage{txfonts}
\usepackage{graphicx}
\begin{document}
\setcounter{page}{187}

   \title{Methanol masers in environments of three massive protostars}

   \author{A. Bartkiewicz \thanks{formerly Niezurawska, e-mail:
          annan@astro.uni.torun.pl}\inst{1},
          M. Szymczak \inst{1},
          \and
          H.J. van Langevelde \inst{2}
          }

   \institute{Toru\'n Centre for Astronomy, Nicolaus Copernicus University,
             Poland
             \and
             Joint Institute for VLBI in Europe, The Netherlands
             }
   \authorrunning{Bartkiewicz et al.}

   \abstract{We present the first EVN maps of 6.7\,GHz methanol masers 
of three high-mass protostar candidates selected from the Toru\'n
unbiased survey of the Galactic plane. A variety of linear and arc like 
structures was detected. A number of maser clusters with projected
sizes of 20$-$100\,AU show monotonic velocity gradients. Some of them
are roughly perpendicular to the major axes of these structures and can arise
behind shock fronts.
   }

   \maketitle

\section{Introduction}
Observations of the 6668.519 MHz methanol maser transition, first detected
by Menten (\cite{menten91}), appear to be powerful tools to
identify the massive early type stars still embedded in their parental dense
molecular clouds. The high brightness of this line enable us to investigate 
structures at milliarcsecond (mas) scales (a few hundreds of AU
at the distances of a few kpc).

Phillips et al. (\cite{phillips98}) analyzed 45 methanol maser sources 
in star-forming regions and divided them into five groups on the basis of 
their morphology: linear (curved), elongated, pair, complex and simple.
The linear masers were outstanding in their survey and showed a monotonic or 
near-monotonic velocity gradient along the source major axis that is 
consistent with a model of masers embedded in a rotating  disk (radius of 
a few thousands of AU) seen edge-on or nearly edge-on around a high mass 
(up to 120 $M_\odot$) protostar or OB star (Norris et al.\,\cite{norris98};
Minier et al.\,\cite{minier00}). 
However, Walsh et al. (\cite{walsh98}) found only 12 sources showing 
velocity gradients in the sample of 97 methanol sites studied.
They proposed a scenario in which the methanol masers appear before
the UC H\,{\small II} phase around the protostar,  associated with embedded
non-ionizing stars. The dense knots of gas are compressed and
accelerated by the passage of the shock and local conditions cause
different geometries of maser spots. 
Dodson et al. (\cite{dodson04}), based on the VLBI data, proposed that
the 6.7\,GHz methanol masers arise behind low-speed ($<$10\,km\,s$^{-1}$) 
planar shock. A shock propagating nearly perpendicular to the line-of-sight 
produces a linear spatial distribution of maser components. 
Interaction of the shock with density perturbations in the star-forming 
region disrupts the linearity of maser structures.

The VLBI results presented below are first in a series for a large sample 
of methanol masers detected in the unbiased survey of the Galactic
plane (Szymczak et al.\ \cite{szymczak02}). Our aims are twofold; 
to investigate the mas scale of 6.7\,GHz methanol maser structures
in order to test the above mentioned hypotheses and to search for
relationships of methanol emission with other tracers of star-forming 
activity. We improved the positions of newly detected 
30 sources in the Toru\'n survey using the single baseline of MERLIN 
(Mark\,II and Cambridge antennas). The absolute positions of the strongest
maser features were determined with accuracies of 0.1$-$0\farcs8 in RA
and 1$-$4\arcsec \, in Dec depending on the source intensity and declination
(Bartkiewicz et al., in prep.).

\section{Observations and data reduction}
    \begin{table}
      \caption[]{Coordinates (J2000) of the target sources.}
         \label{table1}
\begin{tabular}{cllll}
\hline
Source & RA     & Dec        & $\Delta$RA & $\Delta$Dec \\
       & (h m s)& ($^o$ ' " )& (") & (")  \\
\hline
G33.64$-$0.21 & 18 53 32.551 & 00 32 06.525 & 0.3 & 4  \\
G35.79$-$0.17 & 18 57 16.911 & 02 27 52.900 & 0.6 & 3  \\
G36.11$+$0.55 & 18 55 16.814 & 03 05 03.720 & 0.2 & 1.4 \\
            \hline
        \end{tabular}
   \end{table}

The observations of methanol maser emission at 6668.519\,MHz of the three 
star-forming regions G33.64$-$0.21, G35.79$-$0.17 and 
G36.11$+$0.55\footnote{the names of sources follow their galactic 
coordinates} were carried out with the European VLBI Network (EVN) 
on 2003 June 08. Useful data were obtained from four antennas 
(Cambridge, Effelsberg, Jodrell Bank and Onsala). 
In the single dish survey these sources showed complex methanol 
spectra with three of more clearly visible features of flux densities 
higher than 10\,Jy. All of them are OH emitters 
(Szymczak \& G\'erard\ \cite{szymczak04}). The coordinates of 
the target sources and their errors as measured with 
the Mark~II $-$ Cambridge baseline of MERLIN are given in Table 1.

The targets were observed for a total of 12\,hr, together with observations 
of continuum source 3C345 for the purpose of bandpass, delay and rate 
calibration. The source J1907+0127 (0.2\,Jy at 6.7\,GHz) 
was used as a phase calibrator for all three targets. The cycle time between
each target and the phase-calibrator was 5.5\,min $+$ 3.5\,min.
We used a spectral bandwidth of 2\,MHz, resulting in the velocity
coverage of 100\,km\,s$^{-1}$, divided into 1024 channels at correlation
to give a velocity resolution of 0.09\,km\,s$^{-1}$.
The bandwidth was centred at the local standard of rest (LSR) velocity of
65\,km\,s$^{-1}$. Observations were made in left- and right-hand circular
polarization, but since the methanol emission is not strongly circularly
polarized both polarizations were averaged in order to improve 
the signal-to-noise ratio.

The data were correlated on the EVN Mk\,IV Data Processor operated by JIVE. 
We carried out the data reduction with standard procedures for spectra line
observations using the Astronomical Image Processing System ({\small
AIPS}). To make final maps we used 1\,mas pixel separations and an
elliptical restoring beam 14$\times$6\,mas with a PA of $-$1\degr \,
and applied the uniform weighting. The rms noise level was 
of 7$-$10\,mJy beam$^{-1}$ in emission-free Stokes $I$ maps.

We created fringe rate maps of the brightest channels of the targets but
still we failed to find the absolute position of our sources.
The target sources were near zero declination (from $+$0\fdg5 to $+$3\fdg1).
The phase-calibrator J1907+0127, the nearest one from the VLBA Calibrator 
Survey, was about $3\degr$ apart from all three targets. 
Furthermore, due to use of only four EVN telescopes listed above 
the uv-plane coverage was poor for N-S baselines.
It is likely that these factors together with too long phase-referencing 
cycle time precluded a proper phase calibration.
However, the positions of the strongest components of each target 
are known with accuracies better than $0\farcs6$ in RA and $4\arcsec$ 
in Dec (Table 1).

\section{Results and Discussion}

\subsection{G33.64$-$0.21}

   \begin{figure*}
   \centering 
  \vspace{313pt}
\includegraphics{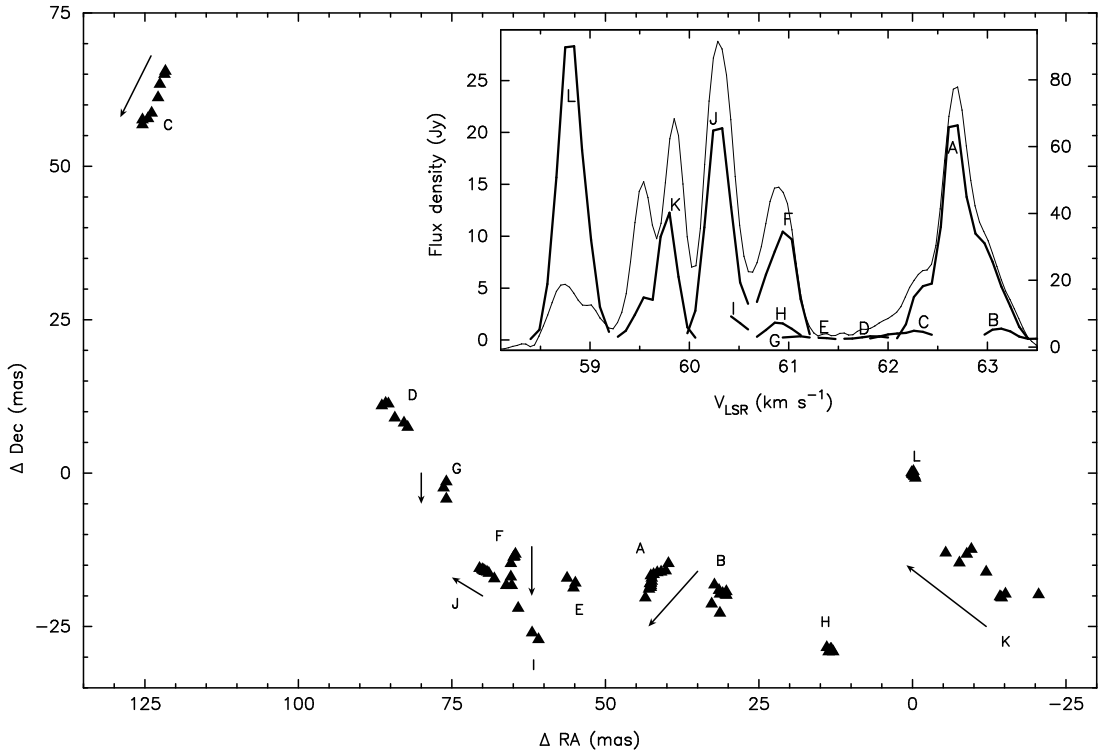}
   \caption{The EVN 6.7\,GHz methanol maser distribution in G33.64$-$0.21. 
The coordinates are relative to the strongest maser component (Table 1).
Each triangle corresponds to a single component found in the channel maps.
Maser clusters A, C, F, G, J and K show clear velocity gradients traced
by arrows (pointed to the blue-shifted velocities). The spectrum (inset) 
is composed of emission profiles of maser clusters labelled from A to L. 
Thin line shows the single dish spectrum taken in 2000 and scaled with 
the right hand ordinate. 
            \label{fig:g33.64}
           }
    \end{figure*}

Fig. 1 shows the spectrum and the overall distribution of maser emission 
in G33.64$-$0.21. The emission is seen in the velocity range 
from 58.5 to 63.3\,km\,s$^{-1}$ and 94 maser components brighter than 
40\,mJy\,b$^{-1}$ (5$\sigma$) are detected. These form 12 clusters
distributed along an arc like structure of overall size of about
150\,mas that corresponds to 585\,AU for the assumed near kinematic
distance of 3.9\,kpc. We do not note any regularity in the velocity 
of the whole structure that can be expected for a rotating disk. 
However, monotonic velocity gradients are found within six clusters named 
as A, C, F, G, J and K (Fig. 1). Angular sizes of individual clusters
range from 5 to 18\,mas i.e. their projected linear sizes range from
20 to 70\,AU. We notice that for most clusters the velocity gradients 
are roughly perpendicular to the arc. 
We state, the methanol maser emission
in G33.64$-$0.21 does not show large-scale velocity gradient but about
half of the maser clusters exhibit internal velocity gradients some of
which are roughly perpendicular to the arc. This can suggest that
the maser structure forms in a shock front postulated by
Dodson et al.\ (\cite{dodson04}). 

The inset in Fig. 1 shows that amplitude of feature L of the EVN spectrum
is comparable (within 30\%) with that of the single dish spectrum. 
For the rest of features the amplitudes observed with the EVN are
a factor of 3-4 lower. This implies, even though variability and/or errors
in flux density cannot be ruled out, that a large fraction 
of maser flux was missed in the interferometric observations. 
Therefore, G33.64$-$0.21 appears as a good candidate to map a low 
intensity and diffuse maser emission.
 
\subsection{G35.79$-$0.17}
   \begin{figure*}
   \centering
   \vspace{250pt}
\includegraphics{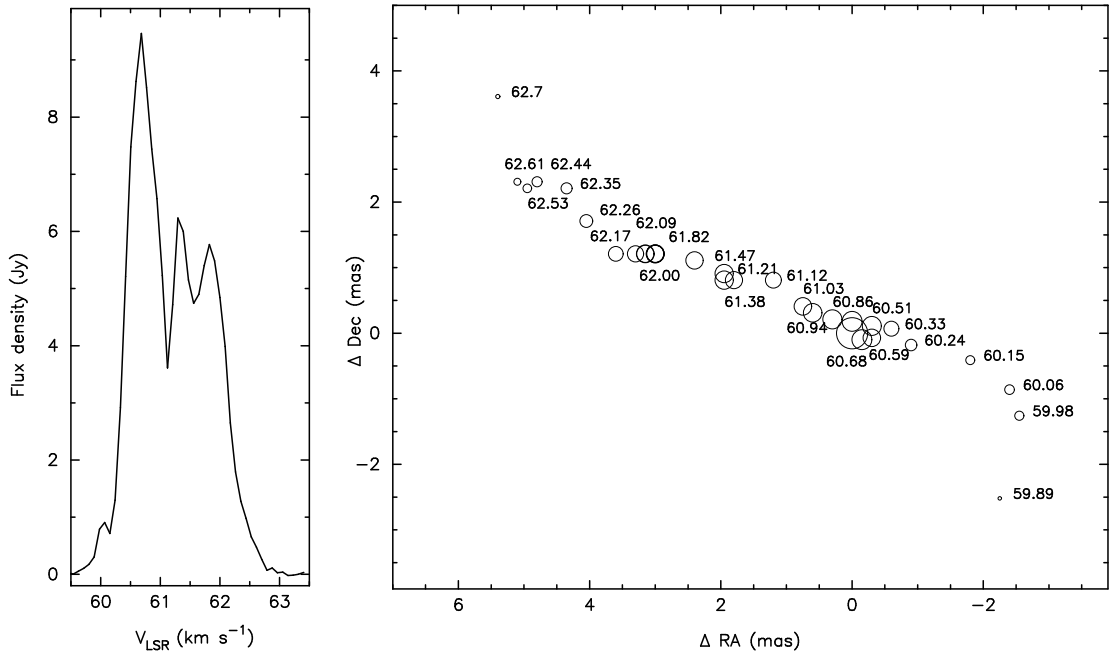}
   \caption{The EVN 6.7\,GHz methanol spectrum (left) and the distribution 
of maser components (right) in G35.79$-$0.17. The coordinates are relative 
to the brightest maser component (Table 1). Each circle corresponds to 
a single component found in the channel maps. The sizes of circles are 
proportional to the logarithm of the brightness. The numbers indicate
the LSR velocities (km\,s$^{-1}$) of the components. 
            \label{fig:g35.79}
           }
    \end{figure*}

The spectrum of G35.79$-$0.17 is composed of four features within 
the velocity range from 59.9 to 62.7\,km\,s$^{-1}$ (Fig. 2). The
shape of the spectrum is the same as observed in 2000 
(Szymczak et al.\,\cite{szymczak02}) but amplitude is about a factor
of 2.5 lower. Variability of the source and calibration errors cannot
fully account for such a difference. We suggest that there is
 diffuse emission missed during our EVN observations.
We found the emission in 33 spectral channels and its distribution  
appears as a 9.5\,mas linear structure (Fig. 2). Analysis of 
the position$-$velocity diagram revealed a linear velocity gradient 
along NE (red-shifted) -- SW (blue-shifted) direction. Such a characteristic 
structure can arise from masers from an edge-on rotating disk 
or accelerating outflow. In a case of an edge-on rotating Keplerian disk 
we can estimate the lower limit of the mass of a central star 
(e.g.\,Phillips et al.\ \cite{phillips98}). Assuming the far and near
kinematic distances of 4.6\,kpc and 10.3\,kpc we infer that the enclosed
mass is 0.4 and 0.9 M$_\odot$, respectively. These values are certainly 
underestimated; we have assumed that there is no inclination and that the
disk size equals the extension of the maser emission. 
Minier et al.\ (\cite{minier00}) found similar structures in W\,51 and 
G\,29.95$-$0.02. They also derived sub-solar masses and concluded from 
a larger sample that they had detected only a small fraction of the disk, 
which typically extends over 1000\,AU. Indeed, the maser structure in 
G35.79$-$0.17 is similar in a size with individual clusters observed
in G33.64$-$0.21 (e.g. the cluster A). We cannot exclude that
the observed structure arises due to an outflow from a massive protostar
signposted by infrared source IRAS18547$+$0223.
Proper motion studies would help to solve the problem of origin of 
the linear structure of methanol maser in G35.79$-$0.17.

\subsection{G36.11$+$0.55}
   \begin{figure*}
  \centering
   \vspace{320pt}
\includegraphics{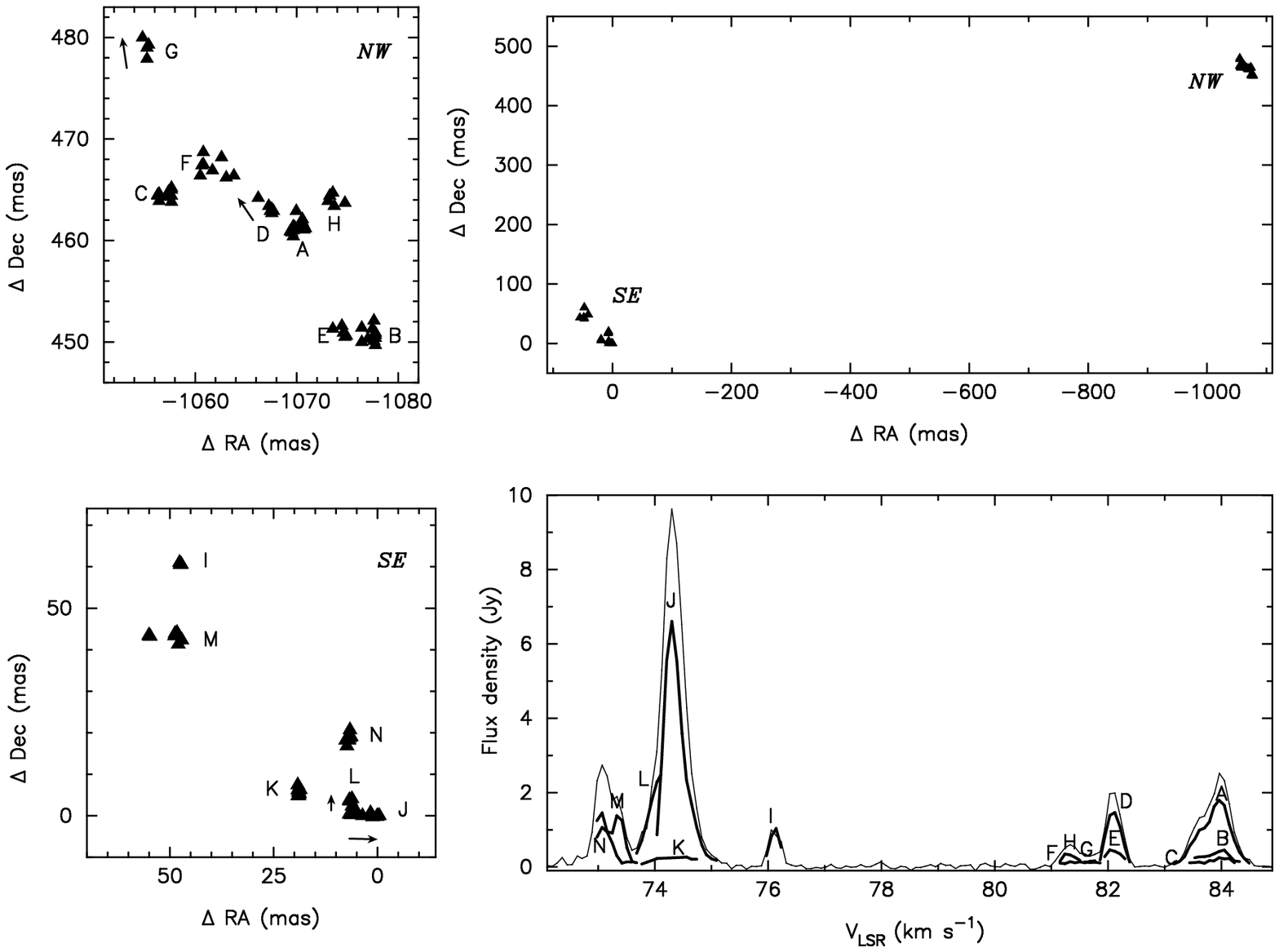}
   \caption{The overall distribution (top right) and spectrum (bottom right)
of the 6.7\,GHz methanol maser emission in G36.11$+$0.55. 
The coordinates are relative to the strongest maser component (Table 1).
Enlarged areas of the NW and SE emission are shown on the left side. 
Each triangle corresponds to a maser component found in the channel maps. 
Arrows mark velocity gradients similarly as in Fig. 1. The spectrum shown
by thin line is the sum of spectra of individual maser clusters (thick line). 
            \label{fig:g36.11}
           }
    \end{figure*}

Fig. 3 shows the spectrum and the distribution of the 6.7\,GHz methanol maser
emission in G36.11$+$0.55. We found two active regions, north-western 
and south-eastern separated by 1\farcs1. The maser components in both regions
are elongated along a position angle of $\sim$45\degr. These structures are
nearly perpendicular to the major axis.  The NW region is composed of
eight clusters, labelled from A to H, within the LSR velocity range from 81.1
to 84.5\,km\,s$^{-1}$. The SE region is composed of six maser clusters, 
labelled from I to N, of the LSR velocity from 73.0 to 76.2\,km\,s$^{-1}$.
4 out of 14 clusters exhibit monotonic velocity gradients; two of them
(NW region) are perpendicular to the major axis. For the assumed near
kinematic distance of 5.3\,kpc the two clumps NW and SE are separated
by 5840\,AU.

The methanol emission of G36.11$+$0.55 lies 5\farcs5 SE from the centre of 
a giant molecular clump Mol\,77 (IRAS18527$+$0301) with embedded high-mass 
protostars of type B2.5$-$O8.5 (Brand et al.\ \cite{brand01}). Since the absolute
position of methanol masers is known within 0\farcs2 and 1\farcs4 (RA and Dec) 
this positional offset may be significant.
The velocities of thermal emission of CS, CO and HCO$^+$ range from
71.9 to 82.1\,km\,s$^{-1}$ with the peaks about 75.5\,km\,s$^{-1}$
and match well the velocity range of methanol emission. 
We note that the overall structure delineated by methanol masers
is nearly perpendicular to the axis of an outflow traced
by the blue- (71.0$-$73.8\,km\,s$^{-1}$) and red-shifted
(79.5$-$83.2\,km\,s$^{-1}$) emission of the $^{13}$CO line as reported
by Brand et al. (\cite{brand01}). Therefore it is very likely that
the 6.7\,GHz maser emission mapped with the EVN are physically related
to a high mass protostar.

\section{Conclusions}
The 6.7\,GHz methanol maser emission towards three star-forming regions 
was imaged with milliarsecond resolution. Linear and arc like structures
were detected. There were considerable numbers of maser clusters with
internal velocity gradients roughly perpendicular to the major axis. 
These geometrical structures can arise in outflows and behind shock fronts.
 
\begin{acknowledgements}
AB acknowledges the support from JIVE during her stay in Dwingeloo.
The European VLBI Network is a joint facility of European, Chinese, 
South African and other radio astronomy institutes funded by their 
national research councils. The work was supported by the KBN grant
2P03D01122.
\end{acknowledgements}

\end{document}